\documentclass[12pt]{article}

\usepackage{scicite}

\usepackage{times}

\usepackage{amsmath}
\usepackage{amssymb}
\usepackage{graphicx}
\usepackage{color}
\usepackage{amsfonts}
\usepackage{subfigure}
\usepackage{enumitem}
\usepackage{url}

\newenvironment{sciabstract}{%
\begin{quote} \bf}
{\end{quote}}

\newcommand{\aAI}{\alpha_{A I}}
\newcommand{\aAR}{\alpha_{A R}}
\newcommand{\aIR}{\alpha_{I R}}
\newcommand{\aEA}{\alpha_{E I_A}}
\newcommand{\aES}{\alpha_{E I_S}}

\newcommand{\Bf}{\mathbf{f}}
\newcommand{\Bg}{\mathbf{g}}
\newcommand{\Bm}{\mathbf{m}}
\newcommand{\Bh}{\mathbf{h}}
\newcommand{\Bu}{\mathbf{u}}

\newcommand{\mi}{\mathbf{m}_i}
\newcommand{\hi}{\mathbf{h}_i}
\newcommand{\ds}{\displaystyle}
\newcommand{\Bup}{\mathbf{u}'}

\newcommand{\td}[2]{\frac{d #1}{d #2}}
\newcommand{\pad}[2]{\frac{\partial #1}{\partial #2}}

\newcommand{\Kt}{\chi} % Number of Covid infection tests per unit time

\definecolor{airforceblue}{rgb}{0.36, 0.54, 0.66}
\definecolor{brickred}{rgb}{0.8, 0.25, 0.33}
\definecolor{applegreen}{rgb}{0.55, 0.71, 0.0}
\definecolor{darkgreen}{rgb}{0.2, 0.5, 0.0}

\newcommand{\luca}[1]{\textcolor{brickred}{\textbf{#1}}}

\title{Lack of practical identifiability may hamper reliable predictions in COVID-19 epidemic models}

\author
{L. Gallo$^{1,2}$, M. Frasca$^3$, V. Latora$^{1,2,4}$, G. Russo$^5$\\
\normalsize{$^{1}$Department of Physics and Astronomy,}\\
\normalsize{University of Catania, 95125 Catania, Italy}\\
\normalsize{$^{2}$INFN Sezione di Catania,}\\
\normalsize{Via S. Sofia, 64, 95125 Catania, Italy}\\
\normalsize{$^3$Department of Electrical, Electronics and Computer Science Engineering,}\\
\normalsize{University of Catania, 95125 Catania, Italy}\\
\normalsize{$^{4}$School of Mathematical Sciences,}\\
\normalsize{Queen Mary University of London, London E1 4NS, UK}\\
\normalsize{$^5$Department of Mathematics and Computer Science,}\\
\normalsize{University of Catania, 95125 Catania, Italy}
}

\date{}

\begin{document} 

% Double-space the manuscript.

\baselineskip24pt

% Make the title.

This paper introduces a general framework to quantify the reliability of the predictions of a dynamical model
in the presence of unmeasurable variables, and presents an application to COVID-19.

\clearpage

\maketitle

% Place your abstract within the special {sciabstract} environment.

\begin{sciabstract}
Compartmental models are widely adopted to describe and predict the spreading of infectious diseases. 
%\vito{@@@} The fitting of the unknown parameters of such models is affected by the uncertainty of the data. 
The unknown parameters of such models need to be estimated from the data.
Furthermore, when some of the model variables are not empirically accessible, as in the case of asymptomatic carriers of COVID-19, they have to be obtained as an outcome of the model. 
%, and are therefore subject to the uncertainty on the parameters.
Here, we introduce a framework to quantify how the uncertainty in the data impacts the determination of the parameters and the evolution of the unmeasured variables of a given model. We illustrate how the method is able to characterize different regimes of identifiability, even in 
models with few compartments.
Finally, we discuss how the lack of identifiability in a realistic model for COVID-19 may prevent reliable forecasting of the epidemic dynamics.
\end{sciabstract}

% In setting up this template for *Science* papers, we've used both
% the \section* command and the \paragraph* command for topical
% divisions.  Which you use will of course depend on the type of paper
% you're writing.  Review Articles tend to have displayed headings, for
% which \section* is more appropriate; Research Articles, when they have
% formal topical divisions at all, tend to signal them with bold text
% that runs into the paragraph, for which \paragraph* is the right
% choice.  Either way, use the asterisk (*) modifier, as shown, to
% suppress numbering.

\section*{Introduction}

The pandemic caused by SARS-CoV-2 is challenging humanity in an unprecedented way \cite{anderson2020will}, with the disease that in a few months has spread around the world affecting large parts of the population \cite{world2020coronavirus,dong2020interactive} and often requiring hospitalization or even intensive care \cite{huang2020clinical,chen2020epidemiological}. Mitigating the impact of COVID-19 urges synergistic efforts to understand, predict and control the many, often elusive, facets of the complex phenomenon of the spreading of a new virus, from RNA sequencing to the study of the virus pathogenicity and transmissibility \cite{wiersinga2020pathophysiology,wang2020review}, to the definition of suitable epidemic spreading models \cite{estrada2020covid} and the investigation of non-pharmaceutical intervention policies and containment measures \cite{chinazzi2020effect,leung2020first,castorina2020data,lanteri2020containment}.
%
%Epidemic spreading models can give an important contribution to these efforts, by providing a better understanding of the mechanisms of disease diffusion and predictive scenarios for the time course of important variables such as number of infected people and patients in intensive care. For  this  reason,  many  models  of  COVID-19  spread-ing have been recently proposed 
In particular, a large number of epidemic models has been recently proposed to  describe  the  evolution  of  COVID-19  and  evaluate  the  effectiveness  of  different  counteracting  measures, including  social  distancing,  testing  and  contact  tracing \cite{fanelli2020analysis,arenas2020mathematical,kucharski2020early,giordano2020modelling,aleta2020modelling,della2020network}. However, even the adoption of well-consolidated modeling techniques, such as the use of mechanistic models at the population level based on compartments, poses fundamental problems. First of all, 
the very same choice of the dynamical variables to use in a compartmental model is crucial, as such variables should adequately capture the spreading mechanisms and need to be tailored to the specific disease. This step is not straightforward, especially when the spreading mechanisms of the disease are still unknown or only partially identified. In addition, some of the variables considered might be difficult to measure and track, as, for instance, it occurs in the case of COVID-19 for the number of individuals showing mild or no symptoms. Secondly, compartmental models, usually, involve a number of parameters, including the initial values of the unmeasured variables, which are not known and need to be estimated from data. 

Having at disposal large amount of data, unfortunately, does not simplify the problem of parameter estimation and prediction of unmeasured states. In fact, once a model is formulated, it may occur that some of its unknown parameters are intrinsically impossible to determine from the measured variables, or that they are numerically very sensitive to the measurements themselves. In the first case, it is the very same structure of the model to hamper parameter estimation, as the system admits infinitely many sets of parameters that fit the data equally well; for this reason, this problem is referred to as \emph{structural identifiability} \cite{heinemann2016model,villaverde2016structural}. In the second case, although, under ideal conditions (i.e., noise-free data and error-free models) the problem of parameter estimation can be uniquely solved, for some trajectories it may be numerically ill-conditioned, such that, from a practical point of view, the parameters cannot be determined with precision even if the model is structurally identifiable. This situation typically occurs when large changes in the parameters entail a small variation of the measured variables, such that two similar trajectories may correspond to very different parameters \cite{quaiser2009systematic}. The term \emph{practical identifiability} is adopted in this case.

Identifiability in general represents an important property of a dynamical system, as in a non-identifiable system different sets of parameters can produce the same or very similar fits of the data. Consequently, predictions from a non-identifiable system become unreliable. 
In the context of epidemics forecasting, this means that even if the model considered is able to reproduce the measured variables, a large uncertainty may affect the estimated values of the parameters and the predicted evolution of the unmeasured variables \cite{roda2020difficult}. 
%
\iffalse
The problem of structural identifiability has been investigated for a large number of COVID-19 epidemic models, showing that these models can have unidentifiable parameters \cite{massonis2020structural}. 
%Conversely, practical identifiability of epidemiological models 
%has being studied in Refs.~\cite{tuncer2018structural,tuncer2018structural2} through 
%Monte Carlo simulations and methods based on information theory, and only to evaluate the sensitivity only of the measured variables on the parameters of the model.
\luca{Moreover, although practical identifiability of epidemiological models has being investigated through simulations and methods based on information theory \cite{tuncer2018structural,tuncer2018structural2}, a detailed analysis of the impact of non-identifiability on the hidden variables is still missing, as only the sensitivity of the measured variables to the parameters of the model is considered}. 
\fi
%
Although the problem of structural identifiability has been investigated already for a large number of COVID-19 epidemic models \cite{massonis2020structural}, the more subtle problem of the practical identifiability of such models has not been faced yet.  
Moreover, in the few existing studies on the  practical identifiability of epidemiological models, only the sensitivity of measured variables to the parameters of the model has been considered, and mainly through numerical simulations \cite{tuncer2018structural,tuncer2018structural2}.
%detailed analysis of the impact of non-identifiability on the hidden variables is still missing 

\medskip
In this paper we investigate the problem of the {\em practical identifiability} of dynamical systems whose state includes not only measurable but also hidden variables, as is the case of compartment models for COVID-19 epidemic. We present a novel and general framework to quantify not only the sensitivity of the measured variables of a given  model on its parameters, but also 
the sensitivity of the unmeasured variables on the 
parameters and on the
measured variables. This will allow us to introduce the notion of {\em practical identifiability of the hidden variables of a model}.
As a relevant and timely application % of our framework 
we show the variety of different regimes and levels of identifiability that can appear in epidemic models, even in the simplest case of a four compartment system. 
Finally, we study the actual effects of the lack of practical identifiability in more  sophisticated models recently introduced for COVID-19.

\section*{Results}
%\subsection*{Modeling and sensitivity measures}
\subsection*{Dynamical systems with hidden variables}
\label{sec:framework}

Consider the $n$-dimensional dynamical system described by the following equations
%\begin{equation}
%\label{eq:general_formM}
%\begin{array}{lll}
%\dot{\mathbf{x}} & = & \mathbf{f}_x(\mathbf{x},\mathbf{y},\mathbf{p}_s),\\
%\dot{\mathbf{y}} & = & \mathbf{f}_y(\mathbf{x},\mathbf{y},\mathbf{p}_s),
%\end{array}
%\end{equation}}
%
\begin{equation}
\label{eq:general_formM}
\begin{array}{lll}
\dot{\mathbf{m}} & = & \mathbf{f}(\mathbf{m},\mathbf{h},\mathbf{q}),\\
\dot{\mathbf{h}} & = & \mathbf{g}(\mathbf{m},\mathbf{h},\mathbf{q}),
\end{array}
\end{equation}
\noindent 
where we have partitioned the state variables into two sets, the variables 
$\mathbf{m}\in \mathbb{R}^{n_m}$ 
that can be empirically accessed (\emph{measurable variables}), 
and those, $\mathbf{h}\in \mathbb{R}^{n_h}$, with $n_m+n_h=n$, that cannot be measured (\emph{hidden}).   
The dynamics of the system is governed by the two Lipschitz-continuous functions 
$\mathbf{f}$ and $\mathbf{g}$, which  
also depend on a vector of structural 
parameters $\mathbf{q}\in \Omega_{q} \subset \mathbb{R}^{n_q}$. The trajectories 
$\mathbf{m}(t)$ and $\mathbf{h}(t)$ 
of system (\ref{eq:general_formM})
are uniquely determined by the structural parameters $\mathbf{q}$ 
and by the initial conditions $\mathbf{m}(0)=\mathbf{m}_0,\mathbf{h}(0)=\mathbf{h}_0$. Here, we assume that some of the quantities $\mathbf{q}$ are known, 
while the others are not known and need to be determined by fitting the trajectories of measurable variables $\mathbf{m}(t)$.
%mathbf{q}
We denote by $\mathbf{p}\in \Omega_{p} \subset \mathbb{R}^{n_p}$ the set of unknown parameters that identify the trajectories, which comprises the unknown terms of $\mathbf{q}$ and the unknown initial conditions $\mathbf{h}_0$. The initial values of the hidden variables are not known, and act indeed as parameters for the trajectories generated by system (\ref{eq:general_formM}). The initial conditions of the measurable variables $\mathbf{m}_0$ may be considered fitting parameters as well.

System (\ref{eq:general_formM}) is said to be {\em structurally identifiable} when the measured variables satisfy \cite{villaverde2016structural}  
\begin{equation}
\mathbf{m}(t,\hat{\mathbf{p}})=\mathbf{m}(t,\mathbf{p}), \forall t \geq 0 \Rightarrow \hat{\mathbf{p}}=\mathbf{p}
\end{equation}
for almost any $\mathbf{p} \in \Omega_p$.
Notice that, as a consequence of the existence and uniqueness theorem for the initial value problem, if system (\ref{eq:general_formM}) 
is structurally identifiable, also the hidden variables can be uniquely determined. 

Structural identifiability guarantees that two different sets of parameters do not lead to the same time course for the measured variables. Clearly, when this condition is not met, one cannot uniquely associate a data fit to a specific set of parameters or, equivalently, recover the parameters from the measured variables \cite{quaiser2009systematic}.

\subsection*{Assessing the practical identifiability of a model}
\label{sec:practical}

Structural  identifiability, however, is a necessary but not sufficient condition for parameters estimation, so that, when it comes to 
use a dynamical system as a model of a real phenomenon, it is fundamental to quantify the  \emph{practical identifiability} of the 
dynamical system. 

To do this, we consider a solution,  $\bar{\mathbf{m}}(t)=\mathbf{m}(t,\bar{\mathbf{p}})$ and $\bar{\mathbf{h}}(t)=\mathbf{h}(t,\bar{\mathbf{p}})$, obtained from parameters $\mathbf{p}=\bar{\mathbf{p}}$, and we explore how much the functions $\mathbf{m}(t)$ and $\mathbf{h}(t)$ change as we vary the parameters $\bar{\mathbf{p}}$ by a small amount $\mathbf{\delta p}$. To first order approximation in the perturbation of the parameters, we have $\mathbf{\delta m}=\frac{\partial \mathbf{m}}{\partial \mathbf{p}}\mathbf{\delta p}+\mathcal{O}(\| \mathbf{\delta p} \|^2)$ and $\mathbf{\delta h}=\frac{\partial \mathbf{h}}{\partial \mathbf{p}}\mathbf{\delta p}+\mathcal{O}(\| \mathbf{\delta p} \|^2)$. 
%\footnote{Here and in the rest of the paper,  $\| \mathbf{v} \|$ denotes the Euclidean norm of a finite dimensional vector $\mathbf{v}$, $\|\mathbf{v}\|^2 = \mathbf{v}\cdot\mathbf{v}$, while for a function $\mathbf{u}(t)$, $\|\mathbf{u}\|$ denotes the $L^2$ norm of $\mathbf{u}$ in $[0,\infty]$, i.e.~$\|\mathbf{u}\|^2 = \int_0^\infty\mathbf{u}\cdot\mathbf{u}\,dt$.}.
Hence, by dropping the higher order terms we have $\|\mathbf{\delta m}\|^2 = \int\limits_0^\infty |\mathbf{\delta m}|^2 dt= \mathbf{\delta p}^T \mathrm{M} \mathbf{\delta p}$ and $\|\mathbf{\delta h}\|^2 = \int\limits_0^\infty |\mathbf{\delta h}|^2 dt= \mathbf{\delta p}^T \mathrm{H} \mathbf{\delta p}$, where the entries of the sensitivity matrices 
$\mathrm{M}=\mathrm{M}(\bar{\mathbf{p}})\in\mathbb{R}^{n_p\times n_p}$ and 
$\mathrm{H}=\mathrm{H}(\bar{\mathbf{p}})\in\mathbb{R}^{n_p\times n_p}$ for the measured and unmeasured variables are defined as
\begin{equation}
\label{eq:sensitivity_matrix}
\mathrm{(M)}_{ij}=\int\limits_0^\infty{\frac{\partial \mathbf{m}^T}{\partial p_i}\frac{\partial \mathbf{m}}{\partial p_j}dt};
\quad
\mathrm{(H)}_{ij}=\int\limits_0^\infty{\frac{\partial \mathbf{h}^T}{\partial p_i}\frac{\partial \mathbf{h}}{\partial p_j}dt}
\end{equation}
Note that these matrices are positive semidefinite by construction.
The smallest change in the measured variables $\mathbf{m}(t)$ will take place if $\mathbf{\delta p}$ is aligned along the eigenvector $\mathbf{v}_1$ of $\mathrm{M}$ corresponding to the smallest eigenvalue $\lambda_1(\mathrm{M})$. Hence, we can consider $\sigma = \sqrt{\lambda_1(\mathrm{M})}$ to quantify 
%how sensitive the solution is to a change in the parameters. 
the \emph{sensitivity of the measured variables to the parameters}.
Practical identifiability requires high values of $\sigma$, as these indicate cases  
where small changes in the parameters may produce considerable variations of the measurable variables, 
and therefore the estimation of the model parameters from fitting is more reliable.

%The minimum eigenvalue $\lambda_1(\mathrm{M}_\mathbf{y})$ thus provides a measure of how sensitive is the solution to a change in the parameters. \luca{Qui dobbiamo scegliere se usare semplicemente $\lambda_1$ oppure se considerare il valore singolare $\sqrt{\lambda_1}$ (io opterei per il primo caso, in verità)} \mattia{DISCUTERE CON GIOVANNI!}

Suppose now we consider a perturbation, $\mathbf{\delta p}_1$, of the parameters aligned along the direction of $\mathbf{v}_1$. We can evaluate the change in $\mathbf{h}(t)$ due to this perturbation by 
\begin{equation}
\eta^2 = \frac{\mathbf{\delta p}_1^T\mathrm{H}\mathbf{\delta p}_1}{\mathbf{\delta p}_1^T\mathbf{\delta p}_1}.
\end{equation}
The value of $\eta$ quantifies the \emph{sensitivity of the hidden variables to the parameters} of the model,  
when such parameters are estimated from the fitting of the observed variables, since $\|\mathbf{\delta h} \|=\eta \|\mathbf{\delta p}_1\|$. 
%{(see Sec.~1 of the SM).}
Notice that in this case and differently from $\sigma$, lower values 
of $\eta$ are desirable because imply a 
better prediction on the hidden variables.

Finally, with the help of the sensitivity matrices defined above, we can also evaluate the \emph{sensitivity 
of the hidden variables to the measured variables} as 
\begin{equation}
\mu^2 = \max\limits_{\|\mathbf{\delta p}\|=1}\frac{\mathbf{\delta p}^T\mathrm{H}\mathbf{\delta p}}{\mathbf{\delta p}^T\mathrm{M}\mathbf{\delta p}}.
\end{equation}

This parameter is of particular relevance here, since it provides a bound on how the uncertainty on the measured variables affects the evolution of the hidden variables. In addition, the parameter $\mu^2$ can be efficiently computed, as it corresponds to the maximum generalized eigenvalue of matrices $(\mathrm{H}$, $\mathrm{M})$, as shown in Methods.

%Note that, if $\mathrm{M}$ is invertible, $\mu^2$ corresponds to the largest eigenvalue of $\mathrm{M}^{-1}\mathrm{H}$.

%\luca{$\mu$ potrebbe essere un ``sensitivity ratio''?}

%The sensitivity measures we have introduced can be used to quantify how the uncertainty of the data affects the estimation of both the parameters and the hidden variables.
%This is, in fact, ultimately determined by the inverse \vito{@@@ ``inverse'' is not clear o lo spieghiamo meglio o leviamo tutta questa seconda frase?} of the relationships discussed above.

The sensitivity matrices are useful in studying the effect of changing the number of hidden variables and unknown parameters on the practical identifiability of a model. Assume we have access to one more variable, thus effectively increasing the size of the set of measured variables to $n_m'=n_m+1$ and, correspondingly, reducing that of the unmeasured variables to $n_h'=n_h-1$. This corresponds to consider new variables $\mathbf{m'}$ and $\mathbf{h'}$.
From the definition in Eq.~(\ref{eq:sensitivity_matrix}), the new sensitivity matrix can be written as $\mathrm{M}^{'}= \mathrm{M} + \mathrm{M}_1$,
%\begin{equation}
%\label{eq:observing_another}
%\begin{array}{lll}
%\mathrm{M}^{'}_{\mathbf{y}} &=& %\mathrm{M}_{\mathbf{y}} + \mathrm{M}_1, \\
%%\mathrm{M}^{'}_{\mathbf{x}} &=& \mathrm{M}_{\mathbf{x}} - \mathrm{M}_1, \\
%\end{array}
%\end{equation}
where $\mathrm{M}_1$ is the sensitivity matrix for the newly measured variable. Given Weyl's inequality (\cite{horn2012matrix}, p.~239) we have that $\lambda_{1}(\mathrm{M}') \geq \lambda_{1}(\mathrm{M}) + \lambda_{1}(\mathrm{M}_1)$
%can write the following relationship for the smallest eigenvalues of the sensitivity matrices
%\begin{equation}
%\label{eq:weyl_inequality}
%\lambda_{1}(\mathrm{M}^{'}_{\mathbf{y}}) \geq \lambda_{1}(\mathrm{M}_{\mathbf{y}}) + \lambda_{1}(\mathrm{M}_1) \geq \lambda_{1}(\mathrm{M}_{\mathbf{y}}),  \end{equation}
and, since $\mathrm{M}_1$ is also positive semi-definite, $\lambda_{1}(\mathrm{M}') \geq \lambda_{1}(\mathrm{M})$. This means that measuring one further variable (or more than one) of the system increases the practical identifiability of a model, as expected. As $\mathrm{H}' = \mathrm{H} - \mathrm{M}_1$, it is also possible to demonstrate that $\mu(\mathrm{M}') \leq \mu(\mathrm{M})$ 
(see Materials and Methods). Let us now consider a different scenario: suppose we have a priori knowledge on one of the model parameters, so that we do not need to estimate its value by fitting the model to the data. In this case, we can define new sensitivity matrices $\widetilde{\mathrm{M}},\widetilde{\mathrm{H}}\in\mathbb{R}^{(n_p-1)\times (n_p-1)}$ for the measured and unmeasured variables respectively. Given the Cauchy's interlacing theorem (\cite{horn2012matrix}, p.~242), we have that $\lambda_{1}(\widetilde{\mathrm{M}}) \geq \lambda_{1}(\mathrm{M})$, 
%Similarly to the previous scenario, as the smallest eigenvalue of $\widetilde{\mathrm{M}}$ is larger than the one of $\mathrm{M}$, we can conclude
%
which implies that practical identifiability is improved by acquiring \textit{a priori} information on some of the model parameters. %For instance, in the context of COVID-19 models, one may decide to determine from fitting the epidemiological parameters that are more difficult to measure, such as the percentage of asymptomatic individuals or the rate of transmission, and to rely on medical and biological knowledge \cite{roda2020difficult,mizumoto2020estimating,bi2020epidemiology,lavezzo2020suppression} to derive the values of other parameters, such as the rate of recovery, which can be therefore assumed to be known in the parameter estimation procedure.
For instance, in the context of COVID-19 models, one may decide to fix some of the parameters, such as the rate of recovery, to values derived from medical and biological knowledge \cite{roda2020difficult,mizumoto2020estimating,bi2020epidemiology,lavezzo2020suppression} and to determine from fitting the more elusive parameters, such as the percentage of asymptomatic individuals or the rates of transmission.

\subsection*{The sensitivity measures reveal different regimes of identifiability}
\begin{figure}[t]
\centering
  \includegraphics[width=0.8\linewidth]{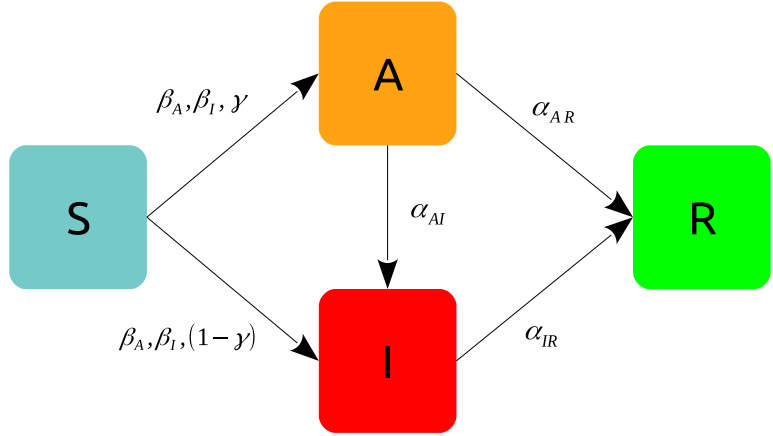}
  \caption{Graphical representation of a SIAR model in which infectious individuals can either be symptomatic (I) or asymptomatic (A)
    (see also Eq.~(\ref{eq:SIAR_model}) in Materials and Methods).}
  \label{fig:SIARmodel}
\end{figure}

As a first application we study the practical identifiability of a four compartment mean-field epidemic model \cite{liu2020new}, in the class of SIAR models \cite{chisholm2018implications},  developed to assess the impact of asymptomatic carriers of COVID-19 \cite{estrada2020covid,pribylova2020seiar,aguilar2020investigating} and other diseases \cite{robinson2013model,balcan2009seasonal,balcan2010modeling}. In such a model (Fig.~\ref{fig:SIARmodel}), a susceptible individual ($S$) can be infected by an infectious individual who can either be symptomatic ($I$) or asymptomatic ($A$). The newly infected individual can either be symptomatic ($S\rightarrow I$) or asymptomatic ($S\rightarrow A$). Furthermore, we also consider the possibility that asymptomatic individuals develop symptoms ($A\rightarrow I$), thus accounting for the cases in which an individual can infect before and after the onset of the symptoms \cite{he2020temporal}. Finally, we suppose that individuals cannot be re-infected, as they acquire a permanent immunity ($R$). 
%Under these assumptions, the equations of the SIAR model are
%\begin{equation}
%\label{eq:SIAR_model}
%\begin{cases}
%\dot{s} = -s(\beta_{I}\iota + \beta_{A}a) \\
%\dot{\iota} = (1- \gamma) s(\beta_{I}\iota + \beta_{A}a) + \alpha_{A I}a - \alpha_{I R}\iota \\
%\dot{a} = \gamma s(\beta_{I}\iota + \beta_{A}a) - (\alpha_{A I} + \alpha_{A R})a \\
%\dot{r} = \alpha_{I R}\iota + \alpha_{A R}a,
%\end{cases}
%\end{equation}

%\noindent where $s(t)$, $\iota(t)$, $a(t)$, $r(t)$ represent the fraction of the population in the four compartments, and $\beta_I$ and $\beta_A$ are the transmission rates for the symptomatic and the asymptomatic individuals respectively, $\gamma$ is the probability for newly infected individuals to show no symptoms, $\alpha_{A I}$ is the rate at which asymptomatic individuals become symptomatic, and $\alpha_{I R}$ and $\alpha_{A R}$ are the recovery rates for the two infectious populations.

One of the crucial aspects of COVID-19 is the presence of  asymptomatic individuals, which are difficult to trace, as the individuals themselves could be unaware about their state. Consequently, we assume that the fraction of asymptomatic individuals, $a(t)$, is not measurable, while the fractions of symptomatic, $\iota(t)$, and recovered, $r(t)$, are measured variables,   
%(at a first approximation, we also assume to be able to trace the asymptomatic individuals once they recover.)
%
that is $\mathbf{m} \equiv [\iota, r]$ and $\mathbf{h} \equiv [s, a]$. 
As mentioned above, practical identifiability is a property of the trajectories of the system, which are uniquely determined by the values of the unknown parameters $\mathbf{p}$. Here, 
%we consider $\mathbf{p}=[a(0), \beta_I, \beta_A, \gamma, \alpha_{A R}]$ and 
we illustrate how the sensitivity of both measured and unmeasured variables change with 
the probability $\gamma$ that a newly infected individual shows no symptoms, when all the other parameters of the model are fixed (to the values reported in Materials and Methods). 

\begin{figure}[t]
\centering
  \includegraphics[width=0.8\linewidth]{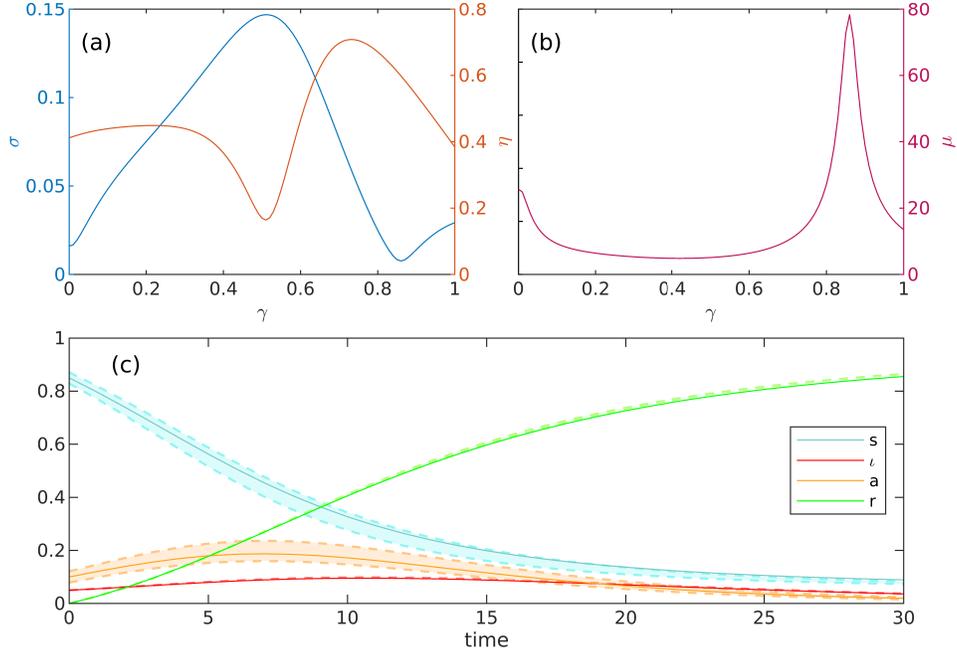}
  \caption{Practical identifiability of the SIAR model in Fig. \ref{fig:SIARmodel}
  %in Eq.~(\ref{eq:SIAR_model})
  as a function of the fraction $\gamma$ of asymptomatic new infectious individuals.
  (a) Sensitivity $\sigma$ and $\eta$ 
   of measured and hidden variables respectively to the parameters of the model. (b) Sensitivity $\mu$ of the hidden variables to the measured ones. (c) State variables for unperturbed values of parameters (with $\gamma = 0.86$, solid line) and for a perturbation with $\|\delta \mathbf{p}\| = 0.3\|\mathbf{p}\|$ along the first eigenvector of $\mathrm{M}$ (dashed lines).}
  \label{fig:findGamma}
\end{figure}

Fig.~\ref{fig:findGamma}(a) shows a non-trivial non-monotonic dependence of our sensitivity measures, $\sigma$ and $\eta$, on $\gamma$. %The specific values of the model parameters, in this case $\gamma$, lead to different levels of identifiability of the measured variables, $I(t)$ and $R(t)$. Similarly, the predictability of the hidden variables, $S(t)$ and $A(t)$, depends on the underlying parameters configuration. 
The value of $\sigma$ has a peak at $\gamma = 0.51$, in correspondence of which $\eta$ takes its minimum value. This represents an optimal condition for practical identifiability,  as the sensitivity to parameters of the measured variables is high, while that of the unmeasured ones is low, 
and this implies that the unknown quantities of the system (both the model parameters and the hidden variables) can be estimated with small uncertainty.
On the contrary, for $\gamma = 0.86$, we observe a relatively small value of $\sigma$ and a large value of $\eta$, meaning that the measured variables are poorly identifiable, and the unmeasured variables are sensitive to a variation of parameters. This is the worst situation in which the  estimated parameters may significantly differ from the real values and the hidden variables may experience large variations even for small changes in the parameters.
%
%In this scenario, we are unable to identify both the parameters determining the dynamics of the measured variables and the time evolution of the hidden variables. 
Furthermore, the quantity $\mu$, which measures the sensitivity of the hidden variables to the measured ones, reported in 
Fig.~\ref{fig:findGamma}(b), exhibits a large peak at the value of $\gamma$ for which $\sigma$ is minimal. This is due to the fact that the vector that determines $\mu$ is almost aligned with $\mathbf{v}_1$. When this holds, we have that $\mu = {\eta}/{\sigma}$, which explains the presence of the spike in the $\mu$ curve. Similarly, the sensitivity $\mu$ takes its minimum almost in correspondence of the maximum of $\sigma$.
The behavior of the model for $\gamma = 0.86$ is further illustrated in Fig.~\ref{fig:findGamma}(c), where the trajectories obtained in correspondence to the unperturbed values of the parameters, i.e., $\mathbf{m}(t,\mathbf{p})$ and $\mathbf{h}(t,\mathbf{p})$ (solid lines), are compared with the dynamics observed 
when $\mathbf{p}$ undergoes a perturbation with $\|\delta \mathbf{p}\| = 0.3\|\mathbf{p}\|$ along $\mathbf{v}_1$ (dashed lines). 
The small sensitivity $\sigma$ of the measured variables $\iota(t,\mathbf{p})$ and $r(t,\mathbf{p})$ to parameters, is reflected into perturbed trajectories that remain close to the unperturbed ones, whereas the large sensitivity $\eta$ of the unmeasured variables $s(t,\mathbf{p})$ and $a(t,\mathbf{p})$ yields perturbed trajectories that significantly deviate from the unperturbed ones.
%The small sensitivity $\sigma$ of the measured variables to parameters, is reflected into perturbed trajectories that remain close to $\mathbf{m}(t,\mathbf{p})$, whereas the large sensitivity of the unmeasured variables, $\eta$, yields perturbed trajectories that significantly deviate from $\mathbf{h}(t,\mathbf{p})$.

\begin{figure}[t]
\centering
  \includegraphics[width=0.8\linewidth]{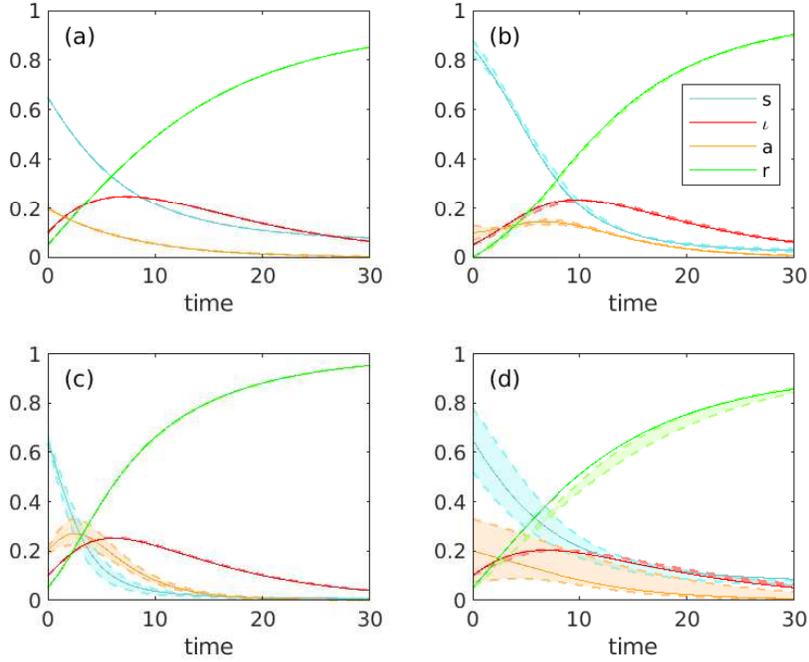}
  \caption{Four scenarios of identifiability for the SIAR model of Fig.~\ref{fig:SIARmodel}.
  %(\ref{eq:SIAR_model}).
  All panels show the system dynamics (solid line) and the evolution of the system when the vector of parameters undergoes a variation $\delta \mathbf{p}$ such that  $\|\delta \mathbf{p}\|= 0.3\|\mathbf{p}\|$ along the first eigenvector of $\mathrm{M}$ (dashed lines). Panels (a) and (c) display configurations for which the observed variables $(\iota, r)$ are not sensitive to the variation, i.e.~the model parameters are not identifiable, while panels (b) and (d) show the opposite case. Furthermore, panels (a) and (b) present scenarios for which the unobserved variables $(s,a)$ are insensitive to the variation, meaning that their are predictable, vice-versa panels (c) and (d) show the the case in which the variables $s$ and $a$ are sensitive.}
  \label{fig:4panels}
\end{figure}

%\luca{Qua inizia cosa c'era nel SM.}

We now illustrate the different levels of identifiability that appears in the SIAR model for diverse settings of the parameters. Its analysis, in fact, fully depicts the more complete perspective on the problem of practical identifiability offered by simultaneously inspecting the sensitivity measures, $\sigma$ and $\eta$. As the two sensitivity measures are not necessarily correlated, there can be cases for which to a high identifiability of the measured variables to the parameters, i.e. large values of $\sigma$, corresponds either a low or a high identifiability of the hidden variables to the parameters. Analogously, for other system configurations, in correspondence of small values of $\sigma$, namely to non-identifiable parameters, one may find large values of $\eta$, meaning that the hidden variables are non-identifiable as well, or, on the contrary small values of $\eta$, indicating that the hidden variables are poorly sensitive to parameter perturbations. Altogether, four distinct scenarios of identifiability can occur and all of them effectively appear in the SIAR model: (a) low identifiability of the model parameters $\mathbf{p}$ and high identifiability of the hidden variables $\mathbf{h}$, (b) high identifiability of both $\mathbf{p}$ and $\mathbf{h}$, (c) low identifiability of both $\mathbf{p}$ and $\mathbf{h}$ and (d) high identifiability of $\mathbf{p}$ and low identifiability of $\mathbf{h}$. To illustrate them, we have considered four distinct configurations of the model (with parameters as given in Table \ref{tab:results_fig_3}, illustrated in Methods) and, for each case study, compared the unperturbed trajectories to the perturbed ones, with the vector of parameters undergoing a variation $\|\delta \mathbf{p}\| = \pm 0.3\|\mathbf{p}\|$, along $\mathbf{v}_1$.
%, namely the eigenvector associated to the smallest eigenvalue of $\mathrm{M}$. 
Fig.~\ref{fig:4panels} shows the results obtained for each parameter configuration. In each panel, the solid lines represent the unperturbed trajectories, while the dashed lines correspond to the perturbed dynamics. {In case (a) and (b), we see that, under the variation $\delta \mathbf{p}$, the perturbed trajectories of the hidden variables remain close to the unperturbed dynamics. Hence, the hidden variables are highly identifiable. Conversely, in case (c) and (d), the perturbed trajectories substantially differ from the unperturbed dynamics, meaning that the hidden variables are poorly identifiable, as they are sensitive to a variation of the model parameters. As concerns the measured variables, in cases (a) and (c) the perturbed trajectories slightly differ from the unperturbed dynamics. Therefore, as the measured variables are insensitive to the perturbation $\delta \mathbf{p}$, the model parameters have a low degree of identifiability. On the other hand, in case (b) and (d), the perturbation of the parameters significantly affects the trajectories of the measured variables, meaning that the set of parameters reproducing the observed data is more identifiable.}

\begin{table}[h!]
    \centering
    \begin{tabular}{c|c|c|c|c}
     & (a) & (b) & (c) & (d) \\
    \hline
    $\sigma$ & $0.0096$& $0.15$ & $0.013$ & $0.091$\\
    \hline
    $\eta$ & $0.012$& $0.16$ & $0.36$ & $1.4$\\
    \hline
    $\mu$ & $34$& $5.2$ & $29$ & $15$\\
    \end{tabular}
    \caption{Values of $\sigma$, $\eta$ and $\mu$ for the four configurations of the SIAR model shown in Fig.~\ref{fig:4panels}.}
    \label{tab:lambda_eta}
\end{table}

{Finally, Table~\ref{tab:lambda_eta} illustrates the values of the sensitivity measures $\sigma$, $\eta$ and $\mu$ for each case. In particular, case (c) represents the worst scenario, as the value of $\sigma$ is relatively small, meaning that the model parameters $\mathbf{p}$ are poorly identifiable, and the value of $\eta$ is large, indicating a high sensitivity of the hidden variables to the parameters. 
Conversely, the best scenario is represented by case (b), for which both the model parameters and the hidden variables are highly identifiable, as the value of $\sigma$ is large compared to the other cases while the value of $\eta$ remains relatively small.}

\subsection*{Lack of identifiability in COVID-19 modeling prevents reliable forecasting}
\begin{figure}[t]
\centering
  \includegraphics[width=0.8\linewidth]{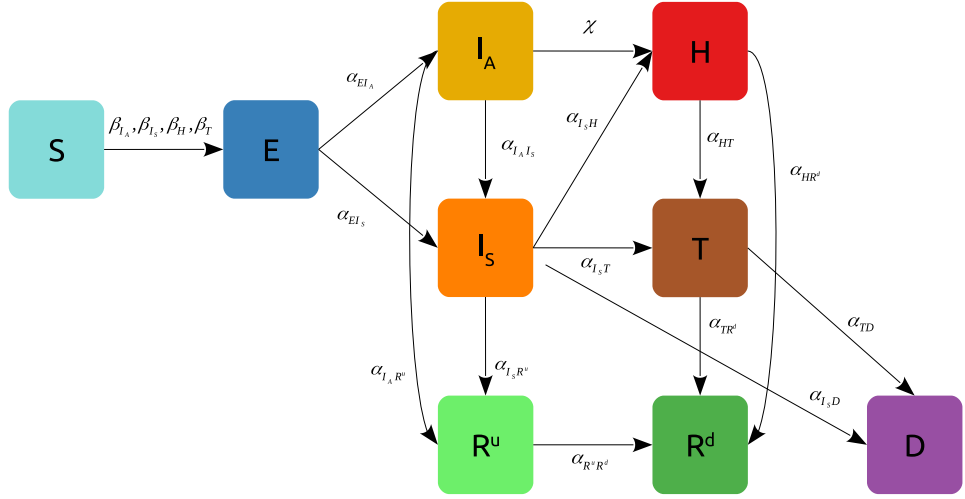}
  \caption{Graphical representation of a nine-compartment model  for the propagation of COVID-19  (see also Eq.~(\ref{eq:var_seidard}) in Materials and Methods).}
  \label{fig:complexSIARmodel}
\end{figure}

\begin{figure}[t!]
\centering
  \includegraphics[width=0.8\linewidth]{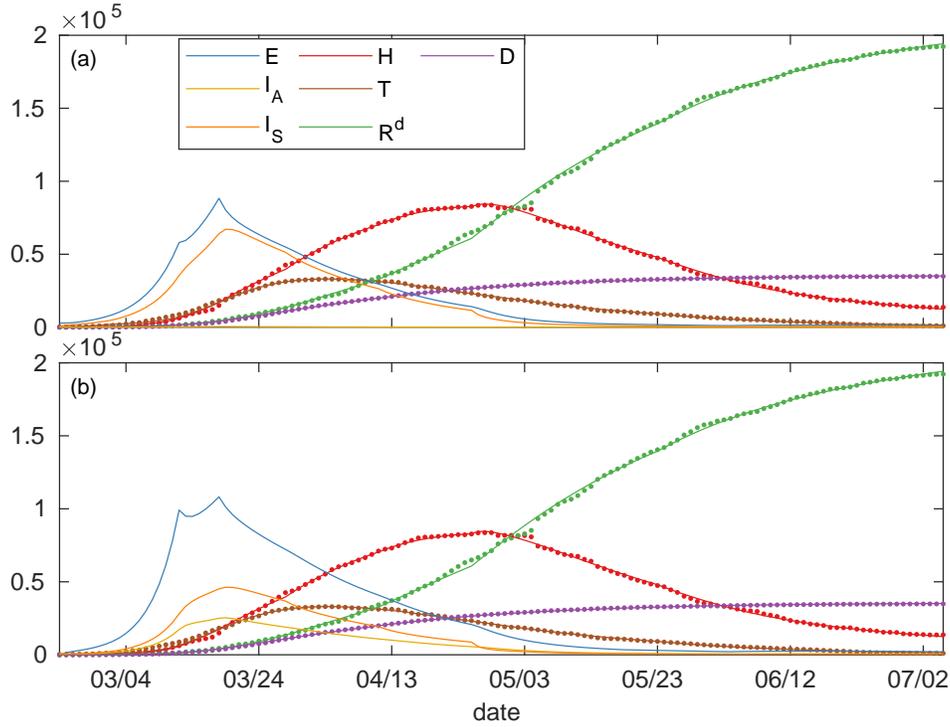}
  \caption{Modelling the 
COVID-19  outbreak  in Italy. The evolution of both hidden and measured variables of the model in Fig. \ref{fig:complexSIARmodel} %Eq.~{(\ref{eq:var_seidard})} 
(continuous lines) is reported together with the official data from the Civil Protection Department (circles). 
(a) and (b) refer to two different sets of parameters $\bf p$ of the model, which fit equally well the available data.}
  \label{fig:2sets}
\end{figure}

As a second application we show the relevance of the problem of practical identifiability in the context of COVID-19 pandemic modeling. We consider a realistic model (Fig.~\ref{fig:complexSIARmodel}) of the disease propagation, that is a variant of the SIDARTHE model \cite{giordano2020modelling} and is characterized by nine compartments accounting respectively for susceptible ($S$), exposed ($E$), undetected asymptomatic ($I_A$), undetected symptomatic ($I_S$), home isolated ($H$), treated in hospital ($T$), undetected recovered ($R^u$), detected recovered ($R^d$) and deceased ($D$) individuals. Following \cite{giordano2020modelling}, in order to account for the different non-pharmaceutical interventions and testing strategies issued during the COVID-19 outbreak in Italy \cite{governpolicies,governlegislation}, the model parameters have been considered piece-wise constant, and estimated, using nonlinear optimization, by fitting of the official data provided by the Civil Protection Department %(Dipartimento della Protezione Civile, 
\cite{pcdata}. 
The available data, namely the daily number of home isolated, hospitalized, detected recovered and deceased individuals, are shown as coloured circles 
in Fig.~\ref{fig:2sets}. 
Accordingly, we have considered four measured and five hidden variables in the model: 
$\mathbf{m} \equiv [H, T, R^d, D]$ and $\mathbf{h} \equiv [S, E, I_A, I_S, R^u]$.
The two panels (a) and (b) of Fig.~\ref{fig:2sets} display the prediction of the model for two sets of the parameters, which have been obtained by allowing different ranges of parameter values for the fitting (see Materials and Methods).
Although the model fits equally well the empirical data on the observed variables in the two cases, fundamental differences in the trend of the hidden variables appear. In particular, in case (a) the model predicts a number of asymptomatic individuals, $I_A$, which is practically zero throughout the entire outbreak, indicating that all the infectious individuals are symptomatic. Instead, a significant population of asymptomatic individuals (25343 in its peak at day 26) 
is observed in case (b) and, correspondingly, the size of the symptomatic population is consistently smaller than in case (a).
%Altogether, these findings lead to the conclusion that the measured variables can be equally fitted by a scenario where asymptomatic individuals are not playing a significant role or, on the contrary, they constitute a large part of the population of infected individuals. Crucially, the 

The two scenarios also lead to remarkable discrepancies in the values of the parameters obtained through fitting. Let us consider, for instance, the rates $\alpha_{E I_A}$ and $\alpha_{E I_S}$, providing information on the percentage of infected individuals not developing symptoms. In scenario (a) $\alpha_{E I_A}=1/18.5$ days$^{-1}$ and $\alpha_{E I_S}=1/2.5$ days$^{-1}$, such that $12\%$ of the newly infected individuals are asymptomatic, while in scenario (b)  $\alpha_{E I_A}=1/7$ days$^{-1}$  and $\alpha_{E I_S}=1/6$ days$^{-1}$, which signifies that only $45\%$ of the individuals do not develop symptoms after the latency period.

These findings have relevant implications. In fact, the large uncertainty on the size of the asymptomatic population makes questionable the use of the model as a tool to decide the policies to adopt, as it is equally consistent with two  scenarios corresponding to two extremely different dynamics of the epidemic.

\section*{Discussion}

The practical identifiability of a dynamical model is a critical, but often neglected, issue in determining the reliability of its predictions. In this paper we have introduced a novel framework to quantify: 
1) the sensitivity of the dynamical variables of a given model to its parameters, even in the presence of variables that are difficult to access empirically; 2) how changes in the measured variables impact the evolution of the unmeasured ones. 
The set of easily computable measures we have introduced enable to assess, for instance, if and when the model predictions on the unmeasured variables are reliable or not, even 
%in the presence of high sensitivity to parameters of the variables used for the fit, i.e. 
when the parameters of the model can be fitted with high accuracy from the available data. 
As we have shown with a series of case studies, practical identifiability can critically affect the predictions of even very refined epidemic models recently introduced for the description of COVID-19, where dynamical variables, such as the population of asymptomatic individuals, are impossible or difficult to measure. 
This, by no means, should question the importance of such models, in that they enable a scenario analysis, otherwise impossible to carry out, and a deeper understanding of the spreading mechanisms of a novel disease, but should hallmark the relevance of a critical analysis of the results that takes into account sensitivity measures. It also highlights the importance of cross-disciplinary efforts that can provide a priori information on some of the parameters, ultimately improving the reliability of a model \cite{estrada2020covid,roda2020difficult}.

\section*{Materials and Methods}
\subsection*{The sensitivity matrices and their properties}
The sensitivity matrices considered in this paper are given by
\begin{equation}
\label{eq:sensitivity_matrix_MM}
\mathrm{M}_{ij}=\int\limits_0^\infty{\frac{\partial \mathbf{m}^T}{\partial p_i}\frac{\partial \mathbf{m}}{\partial p_j}dt};
\quad
\mathrm{H}_{ij}=\int\limits_0^\infty{\frac{\partial \mathbf{h}^T}{\partial p_i}\frac{\partial \mathbf{h}}{\partial p_j}dt},
\end{equation}
where the vector functions $\mathbf{m} = \mathbf{m}(t;\mathbf{p})$ and $\mathbf{h} = \mathbf{h}(t;\mathbf{p})$ are obtained integrating  system (1). 
The derivative of measurable and hidden variables with respect to the parameters $\mathbf{p}$, \[
\mathbf{m}_i\equiv \partial{\mathbf{m}}/\partial p_i, \quad
\mathbf{h}_i\equiv \partial{\mathbf{h}}/\partial p_i
\]
can be obtained by integrating the system
\begin{equation}
\label{eq:time_integration}
\begin{array}{lll}
\ds \td{\mi}{t}  & = & \ds \pad{\Bf}{\Bm}\cdot\mi + \pad{\Bf}{\Bh}\cdot\hi +\pad{\Bf}{p_i},
\\[5mm]
\ds \td{\hi}{t}  & = & \ds \pad{\Bg}{\Bm}\cdot\mi + \pad{\Bg}{\Bh}\cdot\hi + \pad{\Bg}{p_i},
%\frac{\partial \dot{\mathbf{m}}}{\partial p_i} & = & \frac{\partial \mathbf{f}}{\partial %p_i}(\mathbf{m},\mathbf{h},\mathbf{p}),\\
%\frac{\partial \dot{\mathbf{h}}}{\partial p_i} & = & \frac{\partial \mathbf{g}}{\partial %p_i}(\mathbf{m},\mathbf{h},\mathbf{p}),
\end{array}
\end{equation}
where $i = 1,\dots n_p$.

The numerical evaluation of the sensitivity matrices is carried out, first integrating system (\ref{eq:time_integration}) (for this step we use a fourth-order Runge-Kutta solver with adaptive step size control), resampling the trajectories with a sampling period of 1 day, and, then, performing a discrete summation over the sampled trajectories. Moreover, integration is carried out over a finite time interval $[0, \tau]$, with large enough $\tau$. In the context of our work, as we have considered SIR-like epidemic models, we set the value of $\tau$ such that the system has reached a stationary state, i.e. the epidemic outbreak has ended, as every infected individual has eventually recovered (or dead, depending on the model).

We now present an important property of the sensitivity matrices. We will only take into account the set of measured variables $\mathbf{m}$, as similar considerations can be made for the hidden variables. Let us assume to be able to measure only a single variable, so that the vector $\mathbf{m}$ collapses into a scalar function, that we call $m_1(t)$. In this case, the element $\mathrm{M}_{ij}$ of the sensitivity matrix would be simply given by 
\begin{equation}
\label{eq:sensitivity_matrix2}
\mathrm{(M)}_{ij}=\int\limits_0^\infty{\frac{\partial m_1}{\partial p_i}\frac{\partial m_1}{\partial p_j}dt}.
\end{equation}
Let us call this sensitivity matrix $\mathrm{M}_1$.

Consider now a larger set of measured variables $\mathbf{m} = (m_1, m_2, \dots, m_{n_m})$. The quantity $\partial \mathbf{m}^T/\partial p_i \partial \mathbf{m}/\partial p_j$ in Eq.~(\ref{eq:sensitivity_matrix_MM}) is given by
\begin{equation}
\label{eq:sensitivity_matrices_property}
\frac{\partial \mathbf{m}^T}{\partial p_i}\frac{\partial \mathbf{m}}{\partial p_j} = \frac{\partial m_1}{\partial p_i}\frac{\partial m_1}{\partial p_j} + \frac{\partial m_2}{\partial p_i}\frac{\partial m_2}{\partial p_j} + \dots + \frac{\partial m_{n_m}}{\partial p_i}\frac{\partial m_{n_m}}{\partial p_j}.
\end{equation}

Therefore, integrating over time in the interval $[0,\infty)$ and given the linearity property of the integrals, we find that the sensitivity matrix $\mathrm{M}$ of the set of the measured variables is given by the sum of the sensitivity matrices of the single measured variables. Formally we have that
\begin{equation}
\label{eq:linearity_property}
    \mathrm{M} = \mathrm{M}_1 + \mathrm{M}_2 + \dots + \mathrm{M}_{n_m}. 
\end{equation}
This property of the sensitivity matrices is useful to demonstrate how measuring a further variable affects the sensitivity measures $\sigma$ and $\mu$ as discussed in the following subsection and in the Results.  

Finally, because matrices $\mathrm{M}$ and $\mathrm{H}$ are positive semidefinite, their eigenvalues are non-negative. For any positive semidefinite matrix $\mathrm{A}$ of order $m$, we shall denote its  eigenvalues as $0\leq\lambda_1(\mathrm{A})\leq\lambda_2(\mathrm{A})\leq\dots\leq\lambda_m(\mathrm{A})$.

\subsection*{Sensitivity measures and their properties}

Here, we discuss in more detail the sensitivity measures introduced in the Results. First, we want to propose a measure to quantify the practical identifiability of the model parameters given the measured variables. To do this, we need to evaluate the sensitivity of the trajectories of the measured variables to a variation of the model parameters. In fact, if this sensitivity is small, then different sets of parameters will produce very similar trajectories of the measured variables, meaning that the parameters themselves are poorly identifiable. In particular, as a measure of the parameters identifiability, we can consider the worst scenario, namely the case in which the perturbation of the parameters minimizes the change in the measured variables. This happens when the variation of the model parameters $\mathbf{\delta p}$ is aligned along the eigenvector $\mathbf{v}_1$ of $\mathrm{M}$ corresponding to the minimum eigenvalue $\lambda_1(\mathrm{M})$. Given the definition of $\mathrm{M}$ we have that $\|\delta\mathbf{m}\| = \sqrt{\lambda_1(\mathrm{M})}\,\|\delta\mathbf{p}\|$, hence we can consider the quantity
\begin{equation}\label{eq:sigma}
    \sigma = \sqrt{\lambda_1(\mathrm{M})}
\end{equation}
as an estimate of the \emph{sensitivity of the measured variables to the parameters}. Note that, here and in the rest of the paper,  $\| \mathbf{v} \|$ denotes the Euclidean norm of a finite dimensional vector $\mathbf{v}$, $\|\mathbf{v}\|^2 = \mathbf{v}\cdot\mathbf{v}$, while for a function $\mathbf{u}(t)$, $\|\mathbf{u}\|$ denotes the $L^2$ norm of $\mathbf{u}$ in $[0,\infty]$, i.e.~$\|\mathbf{u}\|^2 = \int_0^\infty\mathbf{u}\cdot\mathbf{u}\,dt$.

Let us now focus on the hidden variables $\mathbf{h}$. In general, as the hidden variables are not directly associated to empirical data, the largest uncertainty on the hidden variables is obtained in correspondence of a variation of the parameters along the eigenvector of $\mathrm{H}$ associated to the largest eigenvalue,
%of $\mathrm{H}$
namely $\lambda_{n_p}(\mathrm{H})$.
%Let us now focus on the hidden variables $\mathbf{h}$. In general, as the hidden variables are not constrained to empirical data, the case for which, given a perturbation of the model parameters, the uncertainty on the hidden variables is maximum consists in considering a variation of the parameters along the eigenvector of $\mathrm{H}$ corresponding to the largest eigenvalue $\lambda_{n_p}(\mathrm{H})$.
Hence, to quantify the sensitivity of the hidden variables to the parameters, one may consider 
\begin{equation}\label{eq:eta_max}
    \eta_{MAX} = \sqrt{\lambda_{n_p}(\mathrm{H})}.
\end{equation}
However, it is crucial to note that the hidden variables ultimately depend on the parameters of the model, which are estimated by fitting data that are available for the measured variables only. As a consequence, it is reasonable to consider a quantity that evaluates how the uncertainty on the model parameters (determined by the uncertainty of the measured variables and by their sensitivity to the parameters) affects the identifiability of the hidden variables. Therefore, as a measure of the \emph{sensitivity of the hidden variables to the parameters} we consider
\begin{equation}\label{eq:eta}
    \eta^2 = \frac{\delta\mathbf{p}_1^T\mathrm{H}\delta\mathbf{p}_1}{\delta\mathbf{p}_1^T\delta\mathbf{p}_1},
\end{equation}
where $\delta\mathbf{p}_1$ is a perturbation of the parameters along the eigenvector $\mathbf{v}_1$ of $\mathrm{M}$ corresponding to the minimum eigenvalue $\lambda_1(\mathrm{M})$. Note that, when $\mathbf{v}_1$ and the eigenvector of $\mathrm{H}$ corresponding to the largest eigenvalue $\lambda_{n_p}(\mathrm{H})$ are aligned, by definition we have $\eta = \eta_{MAX}$. 

Finally, we want to define a quantity to estimate how much the hidden variables are perturbed given a variation of the measured ones. In particular, as a measure of the \emph{sensitivity of the hidden variables to the measured variables}, we consider the maximum perturbation of the hidden variables given the minimum variation of the measured ones, which is
\begin{equation}\label{eq:mu}
\mu^2 = \max\limits_{\|\mathbf{\delta p}\|=1}\frac{\mathbf{\delta p}^T\mathrm{H}\mathbf{\delta p}}{\mathbf{\delta p}^T\mathrm{M}\mathbf{\delta p}}.
\end{equation}

Note that $\mu^2$ can be computed considering the following generalized eigenvalue problem
\begin{equation}\label{eq:generalized_eig_prob}
    \mathrm{H}\mathbf{u}_k = \lambda_k\mathrm{M}\mathbf{u}_k,
\end{equation}
where $\mathrm{H}$ and $\mathrm{M}$ are the sensitivity matrices for the hidden and the observed variables respectively, and $\lambda_k = \lambda_k(\mathrm{M},\mathrm{H})$ denotes the $k$-th generalized eigenvalue of matrices $\mathrm{M}$ and $\mathrm{H}$.  
We will denote by $\lambda_{n_p}$ the largest generalized eigenvalue, and $\Bu$ the corresponding generalized eigenvector. Note that, since both matrices are symmetric, if $\mathbf{u}$ is a right eigenvector then $\mathbf{u}^{T}$ is a left eigenvector. Multiplying each member of the equation by $\mathbf{u}^{T}$ and dividing by $\mathbf{u}^{T}\mathrm{M}\mathbf{u}$, we obtain
\begin{equation}\label{eq:sup_def}
    \lambda_{n_p}= \frac{\mathbf{u}^{T}\mathrm{H}\mathbf{u}}{\mathbf{u}^{T}\mathrm{M}\mathbf{u}} = \max\limits_{\|\mathbf{v}\|=1} \frac{\mathbf{v}^{T}\mathrm{H}\mathbf{v}}{\mathbf{v}^{T}\mathrm{M}\mathbf{v}},
\end{equation}
where one can recognize the definition of $\mu^2$ provided in Eq.~(\ref{eq:mu}). In other words, $\mu^2$ represents the largest eigenvalue of the matrix $\mathrm{M}^{-1}\mathrm{H}$. 

It is worth noting two aspects about the sensitivity measure $\mu$. First, given definitions (\ref{eq:sigma}) and (\ref{eq:eta_max}), for any $\delta\mathbf{p}$ with $\|\delta\mathbf{p}\|=1$, we have that $\delta\mathbf{p}^T\mathrm{H}\delta\mathbf{p} \leq \eta_{MAX}^2$ and $\delta\mathbf{p}^T\mathrm{M}\delta\mathbf{p} \geq \sigma^2$. As a consequence, we have that
\begin{equation}
    \mu^2 \leq \frac{\eta_{MAX}^2}{\sigma^2}.
\end{equation}
Second, when the vector $\delta\mathbf{p}$ that determines $\mu$ is aligned with the eigenvector $\mathbf{v}_1$ of $\mathrm{M}$,
it is possible to express $\mu$ in terms of the sensitivity measures $\sigma$ and $\eta$. Indeed, when $\delta\mathbf{p} = \|\delta\mathbf{p}\|\mathbf{v}_1 = \mathbf{v}_1$, recalling definitions (\ref{eq:sigma}) and (\ref{eq:eta}), one obtains $\mathbf{v}_1^T\mathrm{M}\mathbf{v}_1 = \sigma^2$, while $\mathbf{v}_1^T\mathrm{H}\mathbf{v}_1 = \eta^2$, from which it follows
\begin{equation}
    \mu = \frac{\eta}{\sigma}.
\end{equation}
Also, we note that if $\mathbf{v}_1$ and the eigenvector of $\mathrm{H}$ corresponding to its largest eigenvalue are aligned, one obtains that $\mu = \eta_{MAX}/\sigma$, which is the maximum value for the sensitivity measure $\mu$. 

We now demonstrate that the sensitivity of the hidden variables to the measured ones, $\mu^2$, decreases as we measure one further variable. Let us assume now we are able to measure one further variable, thus increasing the size of the set of measured variables to $n_m'=n_m+1$ and, correspondingly, reducing that of the unmeasured variables to $n_h'=n_h-1$. Given the property in Eq.~(\ref{eq:linearity_property}), the new sensitivity matrices can be written as $\mathrm{M}'= \mathrm{M} + \mathrm{M}_1$ and $\mathrm{H}'= \mathrm{H} - \mathrm{M}_1$, where by $\mathrm{M}_1$ we denote the sensitivity matrix for the newly measured variable. The new generalized eigenvalue problem is
\begin{equation}
    \mathrm{H}'\Bup = \lambda'\mathrm{M}'\Bup \Leftrightarrow (\mathrm{H}-\mathrm{M}_1)\Bup = \lambda'(\mathrm{M}+\mathrm{M}_1)\Bup,  
\end{equation}
where, for simplicity, we have denoted by $\lambda'$ the largest generalized eigenvalue of matrices $\mathrm{M}'$ and $\mathrm{H}'$

Left multiplying by $\Bup^{T}$ and dividing by $\Bup^{T}\mathrm{M}\Bup$, we obtain
\begin{equation}
\lambda_{n_p} = \frac{\Bu^T\mathrm{H}\Bu}{\Bu^T\mathrm{M}\Bu} \ge \frac{\Bup^T\mathrm{H}\Bup}{\Bup^T\mathrm{M}\Bup}
 = \frac{\Bup^T \mathrm{H}' \Bup + \Bup^T \mathrm{M}_1 \Bup}{\Bup^T \mathrm{M} \Bup-\Bup^T \mathrm{M}_1\Bup} \ge 
 \frac{\Bup^T \mathrm{H}' \Bup}{\Bup^T \mathrm{M}' \Bup} = \lambda',
\end{equation}
%    \lambda \geq \frac{\mathbf{u}^{'T}\mathrm{H}\mathbf{u}^{'}}{\mathbf{u}^{'T}\mathrm{M}\mathbf{u}^{'}} = \lambda^{'} + (1+\lambda^{'})\frac{\mathbf{u}^{'T}\mathrm{M}_1\mathbf{u}^{'}}{\mathbf{u}^{'T}\mathrm{M}\mathbf{u}^{'}} \geq \lambda^{'},  
where the first inequality comes from the definition of $\lambda_{n_p}$, while the second comes from the fact that $\mathrm{H}$, $\mathrm{M}$, $\mathrm{H}'$, $\mathrm{M}'$ and $\mathrm{M}_1$ are positive semidefinite.  In short, we find that $\lambda_{n_p}\geq\lambda'$, meaning that, by measuring one variable, the sensitivity of the hidden variables to the measured ones decreases.

\subsection*{SIAR model and setup for numerical analysis}

%In this section we further illustrate the SIAR model and we present the setup for the numerical analysis of the different scenarios of identifiability.
%focusing, in particular, on the dependence of the sensitivity measures on the underlying model parameters and the different scenarios arising from considering the identifiability of both the measured and the hidden variables of the system.

The SIAR model of Fig. \ref{fig:SIARmodel} is described by the following equations

\begin{equation}
\label{eq:SIAR_model}
\begin{cases}
\dot{s} = -s(\beta_{I}\iota + \beta_{A}a) \\
\dot{\iota} = (1-\gamma) a(\beta_{I}\iota + \beta_{A}a) + \aAI a - \aIR \iota \\
\dot{a} = \gamma a(\beta_{I}\iota + \beta_{A}a) - (\aAI + \aAR)a \\
\dot{r} = \aIR \iota + \aAR a,
\end{cases}
\end{equation}

\noindent where $s(t)$, $\iota(t)$, $a(t)$, and $r(t)$ represent \emph{population densities}, i.e., $s(t)=S(t)/N$, $\iota(t)=I(t)/N$, $a(t)=A(t)/N$, and $r(t)=R(t)/N$, where $S(t)$, $I(t)$, $A(t)$, and $R(t)$ represent the number of susceptible, infectious, asymptomatic and recovered individuals and $N$ is the size of the population, so that $s(t)+\iota(t)+a(t)+r(t)=1$. Here, $\beta_I$ and $\beta_A$ are the transmission rates for the symptomatic and the asymptomatic individuals respectively, $\gamma$ is the probability for newly infected individuals to show no symptoms, $\alpha_{A I}$ is the rate at which asymptomatic individuals become symptomatic, and $\alpha_{I R}$ and $\alpha_{A R}$ are the recovery rates for the two infectious populations.

Asymptomatic individuals are difficult to trace, as the individuals themselves could be unaware about their state. As a consequence, we assume that the density of asymptomatic individuals is not measurable, while the densities of symptomatic and recovered individuals are measured variables. According to the notation introduced in Eq.~(\ref{eq:general_formM}), we therefore have that $\mathbf{m} \equiv [\iota, r]$ and $\mathbf{h} \equiv [s, a]$. Note that, as a first approximation, here we assume to be able to trace the asymptomatic individuals once they recover.

The results presented in Fig.~\ref{fig:findGamma} have been obtained considering the following setup. As the number of symptomatic infectious and recovered individuals are considered measurable, we have assumed that the initial conditions $\iota(0)$, $r(0)$ and the rate of recovery $\alpha_{I,R}$ are known parameters. Second, we have supposed to be able to measure, for instance through backward contact tracing, the rate at which asymptomatic individuals develop symptoms, i.e., $\alpha_{A I}$. Hence, the vector of parameters to determine by calibrating the model is given by $\mathbf{p}=[a(0), \beta_I, \beta_A, \gamma, \alpha_{A R}]$. Table~\ref{tab:results_fig_2} displays the value of the model parameters used to obtain the results shown in Fig.~\ref{fig:findGamma}.

\begin{table}[h!]
    \centering
    \begin{tabular}{c|c|c||c|c||c|c|c}
     $\iota_0$ & $a_0$ & $r_0$ & $\beta_I$ & $\beta_A$ & $\alpha_{I R}$ & $\alpha_{A R}$ & $\alpha_{A I}$ \\
    \hline
    $0.05$ & $0.1$ & $0$ & $0.6$ & $0.3$ &  $0.1$ & $0.2$ & $0.03$ \\
    \end{tabular}
    \caption{Values of the model parameters used for the case study in Fig.~\ref{fig:findGamma}.}
    \label{tab:results_fig_2}
\end{table}

For the analysis of the four scenarios considered in Fig.~\ref{fig:4panels}, the values of the model parameters have been set as given in Table~\ref{tab:results_fig_3}. Furthermore, to better contrast the results arising in the different case studies, in  (a) and (c) we have considered $\mathbf{p}=[\iota(0), a(0), r(0), \beta_I, \beta_A, \gamma, \alpha_{I R}, \alpha_{A R}, \alpha_{A I}]$, while in (b) and (d) we have set $\mathbf{p}=[a(0), \beta_I, \beta_A, \gamma, \alpha_{A R}]$.

\begin{table}[h!]
    \centering
    \begin{tabular}{c|c|c|c||c|c|c||c|c|c}
     & $\iota_0$ & $a_0$ & $r_0$ & $\beta_I$ & $\beta_A$ & $\gamma$ & $\alpha_{I R}$ & $\alpha_{A R}$ & $\alpha_{A I}$ \\
    \hline
    (a) & $0.1$ & $0.2$ & $0.05$ & $0.3$ & $0.4$ &  $0.26$ & $0.1$ & $0.2$ & $0.03$ \\
    \hline
    (b) & $0.05$  & $0.1$ &  $0$ & $0.6$ & $0.3$ & $0.51$  & $0.1$ & $0.2$ & $0.03$ \\
    \hline 
    (c) & $0.1$ & $0.2$ & $0.05$ & $0.6$  & $0.8$ & $0.77$  &  $0.1$ & $0.2$ & $0.1$ \\
    \hline
    (d) & $0.1$ & $0.2$ & $0.05$ & $0.3$ &  $0.4$ & $0.53$ & $0.1$ & $0.2$ & $0.03$\\
    \end{tabular}
    \caption{Values of the model parameters used for the case study in Fig.~\ref{fig:4panels}}
    \label{tab:results_fig_3}
\end{table}

\subsection*{Nine compartment model for COVID-19}

%In this section, we introduce and analyze a more complex compartmental model, showing how relevant the problem of practical identifiability becomes in the context of COVID-19 pandemic modeling.
The nine compartment model of Fig. \ref{fig:complexSIARmodel} can be considered as a variant of the SIDARTHE model \cite{giordano2020modelling}.
It is characterized by the presence of an incubation state, in which the individuals have been exposed to the virus ($E$) but are not yet infectious, and by infectious individuals, that, in addition to being symptomatic or asymptomatic, can be either detected or undetected. The model, therefore, includes four classes of infectious individuals: undetected asymptomatic ($I_A$), undetected symptomatic and pauci-symptomatic ($I_S$), home isolated ($H$, corresponding to detected asymptomatic and pauci-symptomatic), and treated in hospital ($T$, corresponding to detected symptomatic).  Finally, removed individuals can be undetected ($R^u$), detected ($R^d$) or deceased ($D$).

The model dynamics is described by the following equations
\begin{equation}
\label{eq:var_seidard}
\left \{
\begin{array}{lll}
 \dot{S} & = & - S ( \beta_{I_A} I_A + \beta_{I_S} I_S + \beta_{H} H + \beta_{T} T)/N  \\
 \dot{E}  & = & \> \phantom{ - } \> S ( \beta_{I_A} I_A + \beta_{I_S} I_S + \beta_{H} H + \beta_{T} T)/N \\ 
            &  &  - (\aEA+\aES )E \\
 \dot I_A & = & \> \displaystyle \aEA E - (\alpha_{I_A I_S} +
 \alpha_{I_A R^u} ) I_A - \Kt I_A \\
  \dot{I_S} & = & \> \aES E + \alpha_{I_A I_S}I_A \\
              & &  - (\alpha_{I_S H} + \alpha_{I_S T} + \alpha_{I_S R^u} +   \alpha_{I_S D} ) I_S \\
 \dot{H} & = & \>  \alpha_{I_S H}I_S + \Kt I_A - ( \alpha_{H T} +  \alpha_{H R^d}) H \\
 \dot{T} & = & \>  \alpha_{I_S T}I_S + \alpha_{H T}H - (\alpha_{T,R^d} +  \alpha_{T D}) T \\
 \dot{R}^u  &  = & \>   \alpha_{I_A R^u} I_A   +    \alpha_{I_S R^u} I_S\\
 \dot{R}^d  &   = & \>  \alpha_{H R^d} H     +   \alpha_{T R^d} T\\
 \dot{D} & = & \>  \alpha_{I_S D} I_S     +   \alpha_{T D} T, 
\end{array}
\right.
\end{equation}  
{where the state variables represent the number of individuals in each compartment, $N=60\cdot 10^6$ and $S+E+I_A+I_S+H+T+R^u+R^d+D=N$.}
The official data on the spreading of COVID-19 in Italy made available by the Civil Protection Department (Dipartimento della Protezione Civile, \cite{pcdata}) provide information only on four of the nine compartments of the model, namely the home isolated ($H$), hospitalized ($T$), detected recovered ($R^d$) and deceased individuals ($D$). These compartments constitute the set of the measured variables, while the other variables have to be considered as hidden, that is $\mathbf{m} \equiv [H, T, R^d, D]$ and $\mathbf{h} \equiv [S, E, I_A, I_S, R^u]$.

All the parameters appearing in (\ref{eq:var_seidard}) are considered unknown, thus they need to be determined through fitting the model to the available data. It should also be noted that, as many non-pharma\-ceutical interventions have been issued/lifted and the testing strategy has been changed several times over the course of the epidemics \cite{governpolicies,governlegislation}, not all parameters can be considered constant in the whole period used for the fitting. Hence, similarly to \cite{giordano2020modelling}, we have divided the whole period of investigation (which in our case ranges from February 24 to July 6, 2020) into different windows, within each of which the parameters are assumed to be constant. In each time window, one allows only some parameters to vary, according to what is reasonable to assume will be influenced by the government intervention during that time window.

We distinguish two kinds of events that may require an adaptation of the model parameters. On the one hand, there are the non-pharmaceutical containment policies, aimed at reducing the disease transmission. When such interventions are issued, the value of the parameters $\beta$ may vary. On the other hand, the testing strategy, which affects the probability of detecting infected individuals, was also not uniform in the investigated period. When the testing policy changes, the value of the parameters $\alpha_{I_S H}$, $\alpha_{H T}$ and $\alpha_{H R^d}$ may vary. Here, we notice two important points. First, the value of $\alpha_{I_S T}$ is assumed to be constant in the whole period, as we suppose that there are no changes in how the symptomatic individuals requiring hospitalization are detected. Second, as a change in the sole parameter $\alpha_{I_S H}$ would affect too much the average time an individual remains infected, then also $\alpha_{H T}$ and $\alpha_{H R^d}$ have to be included in the set of parameters that may change. Based on these considerations, the intervals in which each parameter remains constant or may change are identified. This defines the specific piece-wise waveform assumed for each of the parameters appearing in the model and, consequently, the effective number of values that need to be estimated for each parameter. 

Hereafter, we summarize the events defining the different windows in which the whole period of investigation is partitioned:

\begin{enumerate}
    \item On March 2, a policy limiting screening  only to symptomatic individuals is introduced.
    \item On March 12, a partial lockdown is issued.
    \item On March 18, a stricter lockdown, which further limits non-essential activities, is imposed.
    \item On March 28, a wider testing campaign is launched. Starting from this date, as the number of tests has constantly increased, while the number of new infections was decreasing, the parameters are allowed to change every 14 or 28 days, namely on April 11, April 25 and May 23. 
    \item On May 4, a partial lockdown lift is proclaimed.
    \item On May 18, further restrictions are relaxed.
    \item On June 3, inter-regional mobility is allowed. This is the last time the model parameters are changed.
\end{enumerate}

Note that, for the time period until April 5, we have followed the same time partition used in \cite{giordano2020modelling}.

The model parameters have been estimated using a nonlinear optimization procedure (implemented via the function {\verb fmincon } in MATLAB) with the following objective function to minimize

\begin{equation}
\label{eq:fittingerror}
    e=\sqrt{\frac{1}{4\tau}\sum\limits_{k=1}^\tau \left ( (H(k)-\bar{H}(k))^2+(T(k)-\bar{T}(k))^2+(R^d(k)-\bar{R}^d(k))^2+(D(k)-\bar{D}(k))^2\right )}
\end{equation}

\noindent where $\bar{H}(k)$, $\bar{T}(k)$, $\bar{R}^d(k)$, and $\bar{D}(k)$ with $k=1,\ldots,\tau$ ($\tau=134$ days) represent the time series of daily data for isolated, hospitalized, detected recovered and deceased individuals provided by the Civil Protection Department \cite{pcdata}, and ${H}(k)$, ${T}(k)$, $\bar{R}^d(k)$, and $\bar{D}(k)$ are the values of the corresponding variables obtained from the integration of Eqs.~(\ref{eq:var_seidard}). The integration of Eqs.~(\ref{eq:var_seidard}) has been carried out by using a suitable ODE solver with maximum integration step size equal to $10^{-2}$ days and then resampling the data with a sampling period of 1 day. 

Fig.~3 displays two distinct fits of model (\ref{eq:var_seidard}). Here, we provide further details on how they have been obtained. In case b), upper and lower bounds on the parameters $\alpha_{E I_A}$ and $\alpha_{E I_S}$ have been incorporated in the parameter estimation procedure, 
thus constraining the percentage of asymptomatic individuals $p_{A} = \alpha_{E I_A}/(\alpha_{E I_A}+\alpha_{E I_S})$, 
while in case (a) no constraint has been considered.  In more detail, 
 in case (b), we fixed $30\% \leq p_A \leq 50\%$ \cite{lavezzo2020suppression} and also imposed $\alpha_{E I_S}/2 \leq \alpha_{E I_A} \leq \alpha_{E I_S}$. In both cases the model fits well the empirical data on the observed variables [the fitting error (\ref{eq:fittingerror}) is $e=898$ individuals in case (a) and $e=938$ individuals in case (b)], but fundamental differences in the trend of the hidden variables appear. In particular, in scenario (a), which does not include constraints on the percentage of asymptomatic individuals, the compartment $I_A$ is approximately zero throughout the entire epidemics, indicating that all the undetected infectious individuals are symptomatic. Vice-versa, in scenario (b) a number of undetected asymptomatic individuals appears, and, correspondingly, the population of undetected symptomatic individuals is consistently smaller than in case (a).

%\{VERSIONE 1 (corta):} The two scenarios also lead to remarkable discrepancies in the values of the parameters obtained through fitting (see for instance the values of the rates $\alpha_{E,A}$ and $\alpha_{E,I}$, as discussed in SM). 

%The two scenarios also lead to remarkable discrepancies in the values of the parameters obtained through fitting. Let us consider, for instance, the rates $\alpha_{E I_A}$ and $\alpha_{E I_S}$, providing information on the percentage of infected individuals not developing symptoms. In scenario a) $\alpha_{E I_A}=0.054$ days$^{-1}$ and $\alpha_{E I_S}=0.40$ days$^{-1}$, such that $12\%$ of the newly infected individuals are asymptomatic, while in scenario b) $\alpha_{E I_A}=0.14$ days$^{-1}$  and $\alpha_{E I_S}=0.17$ days$^{-1}$, which signifies that only $45\%$ of the individuals do not develop symptoms after the latency period. 

% Your references go at the end of the main text, and before the
% figures.  For this document we've used BibTeX, the .bib file
% scibib.bib, and the .bst file Science.bst.  The package scicite.sty
% was included to format the reference numbers according to *Science*
% style.

%BibTeX users: After compilation, comment out the following two lines and paste in
% the generated .bbl file. 

\bibliography{biblio}

\begin{thebibliography}{10}

\bibitem{anderson2020will}
R.~M. Anderson, H.~Heesterbeek, D.~Klinkenberg, T.~D. Hollingsworth, How will
  country-based mitigation measures influence the course of the covid-19
  epidemic?
\newblock {\it The Lancet\/} {\bf 395}, 931--934 (2020).

\bibitem{world2020coronavirus}
{World Health Organization (WHO)}, Coronavirus disease (covid-19): weekly
  epidemiological update,
  \url{https://www.who.int/emergencies/diseases/novel-coronavirus-2019/situation-reports}
  (2020). Accessed: 2020-10-15.

\bibitem{dong2020interactive}
E.~Dong, H.~Du, L.~Gardner, An interactive web-based dashboard to track
  covid-19 in real time.
\newblock {\it The Lancet infectious diseases\/} {\bf 20}, 533--534 (2020).

\bibitem{huang2020clinical}
C.~Huang, Y.~Wang, X.~Li, L.~Ren, J.~Zhao, Y.~Hu, L.~Zhang, G.~Fan, J.~Xu,
  X.~Gu, {\it et~al.\/}, Clinical features of patients infected with 2019 novel
  coronavirus in wuhan, china.
\newblock {\it The lancet\/} {\bf 395}, 497--506 (2020).

\bibitem{chen2020epidemiological}
N.~Chen, M.~Zhou, X.~Dong, J.~Qu, F.~Gong, Y.~Han, Y.~Qiu, J.~Wang, Y.~Liu,
  Y.~Wei, {\it et~al.\/}, Epidemiological and clinical characteristics of 99
  cases of 2019 novel coronavirus pneumonia in wuhan, china: a descriptive
  study.
\newblock {\it The Lancet\/} {\bf 395}, 507--513 (2020).

\bibitem{wiersinga2020pathophysiology}
W.~J. Wiersinga, A.~Rhodes, A.~C. Cheng, S.~J. Peacock, H.~C. Prescott,
  Pathophysiology, transmission, diagnosis, and treatment of coronavirus
  disease 2019 (covid-19): a review.
\newblock {\it Jama\/} {\bf 324}, 782--793 (2020).

\bibitem{wang2020review}
L.-s. Wang, Y.-r. Wang, D.-w. Ye, Q.-q. Liu, A review of the 2019 novel
  coronavirus (covid-19) based on current evidence.
\newblock {\it International journal of antimicrobial agents\/} p. 105948
  (2020).

\bibitem{estrada2020covid}
E.~Estrada, Covid-19 and sars-cov-2. modeling the present, looking at the
  future.
\newblock {\it Physics Reports\/}  (2020).

\bibitem{chinazzi2020effect}
M.~Chinazzi, J.~T. Davis, M.~Ajelli, C.~Gioannini, M.~Litvinova, S.~Merler,
  A.~P. y~Piontti, K.~Mu, L.~Rossi, K.~Sun, {\it et~al.\/}, The effect of
  travel restrictions on the spread of the 2019 novel coronavirus (covid-19)
  outbreak.
\newblock {\it Science\/} {\bf 368}, 395--400 (2020).

\bibitem{leung2020first}
K.~Leung, J.~T. Wu, D.~Liu, G.~M. Leung, First-wave covid-19 transmissibility
  and severity in china outside hubei after control measures, and second-wave
  scenario planning: a modelling impact assessment.
\newblock {\it The Lancet\/}  (2020).

\bibitem{castorina2020data}
P.~Castorina, A.~Iorio, D.~Lanteri, Data analysis on coronavirus spreading by
  macroscopic growth laws.
\newblock {\it International Journal of Modern Physics C\/} p. 2050103 (2020).

\bibitem{lanteri2020containment}
D.~Lanteri, D.~Carc{\`o}, P.~Castorina, M.~Ceccarelli, B.~Cacopardo,
  Containment effort reduction and regrowth patterns of the covid-19 spreading.
\newblock {\it arXiv preprint arXiv:2004.14701\/}  (2020).

\bibitem{fanelli2020analysis}
D.~Fanelli, F.~Piazza, Analysis and forecast of covid-19 spreading in china,
  italy and france.
\newblock {\it Chaos, Solitons \& Fractals\/} {\bf 134}, 109761 (2020).

\bibitem{arenas2020mathematical}
A.~Arenas, W.~Cota, J.~Gomez-Gardenes, S.~G{\'o}mez, C.~Granell, J.~T.
  Matamalas, D.~Soriano-Panos, B.~Steinegger, A mathematical model for the
  spatiotemporal epidemic spreading of covid19.
\newblock {\it MedRxiv\/}  (2020).

\bibitem{kucharski2020early}
A.~J. Kucharski, T.~W. Russell, C.~Diamond, Y.~Liu, J.~Edmunds, S.~Funk, R.~M.
  Eggo, F.~Sun, M.~Jit, J.~D. Munday, {\it et~al.\/}, Early dynamics of
  transmission and control of covid-19: a mathematical modelling study.
\newblock {\it The lancet infectious diseases\/}  (2020).

\bibitem{giordano2020modelling}
G.~Giordano, F.~Blanchini, R.~Bruno, P.~Colaneri, A.~Di~Filippo, A.~Di~Matteo,
  M.~Colaneri, Modelling the covid-19 epidemic and implementation of
  population-wide interventions in italy.
\newblock {\it Nature Medicine\/} pp. 1--6 (2020).

\bibitem{aleta2020modelling}
A.~Aleta, D.~Mart{\'\i}n-Corral, A.~P. y~Piontti, M.~Ajelli, M.~Litvinova,
  M.~Chinazzi, N.~E. Dean, M.~E. Halloran, I.~M. Longini~Jr, S.~Merler, {\it
  et~al.\/}, Modelling the impact of testing, contact tracing and household
  quarantine on second waves of covid-19.
\newblock {\it Nature Human Behaviour\/} {\bf 4}, 964--971 (2020).

\bibitem{della2020network}
F.~Della~Rossa, D.~Salzano, A.~Di~Meglio, F.~De~Lellis, M.~Coraggio,
  C.~Calabrese, A.~Guarino, R.~Cardona-Rivera, P.~De~Lellis, D.~Liuzza, {\it
  et~al.\/}, A network model of italy shows that intermittent regional
  strategies can alleviate the covid-19 epidemic.
\newblock {\it Nature Communications\/} {\bf 11}, 1--9 (2020).

\bibitem{heinemann2016model}
T.~Heinemann, A.~Raue, Model calibration and uncertainty analysis in signaling
  networks.
\newblock {\it Current opinion in biotechnology\/} {\bf 39}, 143--149 (2016).

\bibitem{villaverde2016structural}
A.~F. Villaverde, A.~Barreiro, A.~Papachristodoulou, Structural identifiability
  of dynamic systems biology models.
\newblock {\it PLoS computational biology\/} {\bf 12}, e1005153 (2016).

\bibitem{quaiser2009systematic}
T.~Quaiser, M.~M{\"o}nnigmann, Systematic identifiability testing for
  unambiguous mechanistic modeling--application to jak-stat, map kinase, and
  nf-$\kappa$ b signaling pathway models.
\newblock {\it BMC systems biology\/} {\bf 3}, 50 (2009).

\bibitem{roda2020difficult}
W.~C. Roda, M.~B. Varughese, D.~Han, M.~Y. Li, Why is it difficult to
  accurately predict the covid-19 epidemic?
\newblock {\it Infectious Disease Modelling\/}  (2020).

\bibitem{massonis2020structural}
G.~Massonis, J.~R. Banga, A.~F. Villaverde, Structural identifiability and
  observability of compartmental models of the covid-19 pandemic.
\newblock {\it arXiv preprint arXiv:2006.14295\/}  (2020).

\bibitem{tuncer2018structural}
N.~Tuncer, T.~T. Le, Structural and practical identifiability analysis of
  outbreak models.
\newblock {\it Mathematical biosciences\/} {\bf 299}, 1--18 (2018).

\bibitem{tuncer2018structural2}
N.~Tuncer, M.~Marctheva, B.~LaBarre, S.~Payoute, Structural and practical
  identifiability analysis of zika epidemiological models.
\newblock {\it Bulletin of mathematical biology\/} {\bf 80}, 2209--2241 (2018).

\bibitem{horn2012matrix}
R.~A. Horn, C.~R. Johnson, {\it Matrix analysis\/} (Cambridge university press,
  2012).

\bibitem{mizumoto2020estimating}
K.~Mizumoto, K.~Kagaya, A.~Zarebski, G.~Chowell, Estimating the asymptomatic
  proportion of coronavirus disease 2019 (covid-19) cases on board the diamond
  princess cruise ship, yokohama, japan, 2020.
\newblock {\it Eurosurveillance\/} {\bf 25}, 2000180 (2020).

\bibitem{bi2020epidemiology}
Q.~Bi, Y.~Wu, S.~Mei, C.~Ye, X.~Zou, Z.~Zhang, X.~Liu, L.~Wei, S.~A. Truelove,
  T.~Zhang, {\it et~al.\/}, Epidemiology and transmission of covid-19 in
  shenzhen china: Analysis of 391 cases and 1,286 of their close contacts.
\newblock {\it MedRxiv\/}  (2020).

\bibitem{lavezzo2020suppression}
E.~Lavezzo, E.~Franchin, C.~Ciavarella, G.~Cuomo-Dannenburg, L.~Barzon,
  C.~Del~Vecchio, L.~Rossi, R.~Manganelli, A.~Loregian, N.~Navarin, {\it
  et~al.\/}, Suppression of covid-19 outbreak in the municipality of vo, italy.
\newblock {\it medRxiv\/}  (2020).

\bibitem{liu2020new}
C.~Liu, X.~Wu, R.~Niu, X.~Wu, R.~Fan, A new sair model on complex networks for
  analysing the 2019 novel coronavirus (covid-19).
\newblock {\it Nonlinear Dynamics\/} pp. 1--11 (2020).

\bibitem{chisholm2018implications}
R.~H. Chisholm, P.~T. Campbell, Y.~Wu, S.~Y. Tong, J.~McVernon, N.~Geard,
  Implications of asymptomatic carriers for infectious disease transmission and
  control.
\newblock {\it Royal Society open science\/} {\bf 5}, 172341 (2018).

\bibitem{pribylova2020seiar}
L.~Pribylova, V.~Hajnova, Seiar model with asymptomatic cohort and consequences
  to efficiency of quarantine government measures in covid-19 epidemic.
\newblock {\it arXiv preprint arXiv:2004.02601\/}  (2020).

\bibitem{aguilar2020investigating}
J.~B. Aguilar, J.~S. Faust, L.~M. Westafer, J.~B. Gutierrez, Investigating the
  impact of asymptomatic carriers on covid-19 transmission.
\newblock {\it medRxiv\/}  (2020).

\bibitem{robinson2013model}
M.~Robinson, N.~I. Stilianakis, A model for the emergence of drug resistance in
  the presence of asymptomatic infections.
\newblock {\it Mathematical biosciences\/} {\bf 243}, 163--177 (2013).

\bibitem{balcan2009seasonal}
D.~Balcan, H.~Hu, B.~Goncalves, P.~Bajardi, C.~Poletto, J.~J. Ramasco,
  D.~Paolotti, N.~Perra, M.~Tizzoni, W.~Van~den Broeck, {\it et~al.\/},
  Seasonal transmission potential and activity peaks of the new influenza a
  (h1n1): a monte carlo likelihood analysis based on human mobility.
\newblock {\it BMC medicine\/} {\bf 7}, 45 (2009).

\bibitem{balcan2010modeling}
D.~Balcan, B.~Gon{\c{c}}alves, H.~Hu, J.~J. Ramasco, V.~Colizza, A.~Vespignani,
  Modeling the spatial spread of infectious diseases: The global epidemic and
  mobility computational model.
\newblock {\it Journal of computational science\/} {\bf 1}, 132--145 (2010).

\bibitem{he2020temporal}
X.~He, E.~H. Lau, P.~Wu, X.~Deng, J.~Wang, X.~Hao, Y.~C. Lau, J.~Y. Wong,
  Y.~Guan, X.~Tan, {\it et~al.\/}, Temporal dynamics in viral shedding and
  transmissibility of covid-19.
\newblock {\it Nature medicine\/} {\bf 26}, 672--675 (2020).

\bibitem{governpolicies}
{Presidenza del Consiglio dei Ministri (Presidency of the Council of
  Ministers)}, Governmental containment policies,
  \url{http://www.governo.it/it/coronavirus-misure-del-governo}. Accessed:
  2020-07-06.

\bibitem{governlegislation}
{Presidenza del Consiglio dei Ministri (Presidency of the Council of
  Ministers)}, Legislation issued in response to covid-19 epidemic,
  \url{http://www.governo.it/it/coronavirus-normativa}. Accessed: 2020-07-06.

\bibitem{pcdata}
{Dipartimento della Protezione Civile (Civil protection department)}, Data on
  the national trend,
  \url{https://github.com/pcm-dpc/COVID-19/tree/master/dati-andamento-nazionale}.
  Accessed: 2020-07-06.

\end{thebibliography}

\bibliographystyle{ScienceAdvances}

\section*{Acknowledgements}

The authors would like to thank Prof.~Valeria Simoncini for pointing out the relation between the sensitivity $\mu$ and the  generalized eigenvalue.

V.L. acknowledges support from the Leverhulme Trust Research Fellowship 278 ``CREATE: The network components of creativity and success''.

V.L and G.R. acknowledge support from University of Catania project ``Piano della Ricerca 2020/2022, Linea d’intervento 2, MOSCOVID''.

\section*{Author contributions}  
L.G., M.F., V.L. and G.R. conceived the research and developed the theory. L.G. carried out the numerical analysis. All authors wrote the manuscript.

\section*{Competing interests} The authors declare they have no competing interests.

\section*{Data and material availability} All data needed to evaluate the results are present in the paper itself. Additional data related to this paper may be requested to the authors.
%\clearpage

\end{document}